\documentclass[sigconf]{acmart}
\setcopyright{rightsretained}
\copyrightyear{2026}
\acmYear{2026}
\acmConference[GRAIL @ CIKM '26]{GRAIL 2026: Workshop on Generative, Retrieval-augmented, and Agentic Intelligence for Personalization}{November 8, 2026}{Rome, Italy}
\acmDOI{}
\acmISBN{}

\usepackage{tikz}
\usetikzlibrary{arrows.meta, positioning, calc, fit}
\usepackage{float}   %
\usepackage{dblfloatfix}  %

\begin{document}
\title{PALATE: Pretrained Autoregressive LATent Embeddings}
\author{Sizhang Zhao}
\email{sizhang.zhao@doordash.com}
\affiliation{%
  \institution{DoorDash}
  \city{Seattle}
  \state{WA}
  \country{USA}
}

\author{Praveen Gondhi}
\email{praveen.gondhi@doordash.com}
\affiliation{%
  \institution{DoorDash}
  \city{Seattle}
  \state{WA}
  \country{USA}
}

\author{Bin Li}
\email{bin.li1@doordash.com}
\affiliation{%
  \institution{DoorDash}
  \city{San Francisco}
  \state{CA}
  \country{USA}
}

\author{Utsaw Kumar}
\email{utsaw.kumar@doordash.com}
\affiliation{%
  \institution{DoorDash}
  \city{San Francisco}
  \state{CA}
  \country{USA}
}
\renewcommand{\shortauthors}{Zhao, Gondhi, Li \& Kumar}

\begin{abstract}
Local-commerce and food-delivery marketplaces run large ads businesses: across
hundreds of thousands of stores, merchants pay to be surfaced on browse and
search surfaces, and a ranker decides which ads each consumer sees. Ranking well demands strong per-consumer personalization, yet a
consumer's raw interaction history is sparse and high-cardinality. Our
production ranker had personalized only through hand-engineered signals: aggregate engagement counts, categorical embeddings, and order-agnostic pooled
history --- with no learned per-consumer sequence representation. We learn one
by training a GPT-style generative model of the consumer's engagement
sequence to predict their next few engagements, and read its last-token hidden
state as the consumer embedding. That single frozen vector becomes one shared
personalization input for every ad surface: a daily job refreshes it, and each
ranker consumes it as an ordinary dense feature. Offline it is the largest
single-feature gain in our ranking-AUC ladder, a relative Eval-AUC gain of
+0.59\% on a browse-driven feed and +0.57\% on a query-driven search surface. Online, it shipped as one
component of bundled ranker retrains on both surfaces: Discovery (revenue +0.79\%, conversion rate +1.17\%) and Search (revenue +0.94\%, conversion rate +1.50\%). Three design choices adapt the generative objective to a repurchase-heavy
marketplace, and are the core of our contribution: an exploration-weighted loss
that up-weights businesses new to the consumer over ones already engaged; a
two-stage curriculum: a click pretrain that keeps the representation broad and
exploratory, then a conversion fine-tune that sharpens purchase intent; and a
designed negative set that pairs a fixed popular-business pool and businesses
shown but not ordered with random and in-batch negatives.

\end{abstract}


\maketitle

\section{Introduction}
\label{sec:intro}

DoorDash is a local-commerce marketplace where consumers browse and order from
hundreds of thousands of stores across food and other verticals. A large part of
how those stores get discovered is through ads: merchants pay to be
surfaced on Discovery (the homepage carousel and homefeed) and in Search results, and a
ranker decides which ads each consumer sees. That
ranking is only as good as the personalization behind it: the right store for
one consumer is the wrong store for another.

The natural signal for that intent is the consumer's own interaction history,
but it is sparse and high-cardinality: each consumer touches a tiny slice of a
store space with hundreds of thousands of entries. Engagement is also
repurchase-heavy --- most consumers re-order from a handful of favorites --- so
the history is dominated by repeats and the harder problem is surfacing new
stores a consumer would try. For personalization, our production ranker relied only on hand-engineered
signals: hundreds of aggregate click, order, and conversion counts and rates over
fixed windows; sparse consumer and context categoricals; and recently engaged
store and cuisine IDs, mean-pooled into order-agnostic bags. The
gap was a learned per-consumer sequence representation. We instead model the
history generatively: we train a GPT-style model on a consumer's
store-engagement sequence to predict the next few businesses, and hand its
last-token hidden state to a DCNv2 ranker as a frozen, fixed-length embedding.

We train the model in two stages, moving from clicks to conversions, under an
exploration-weighted loss that up-weights new businesses over repeats to counter
the repurchase-heavy domain.
Offline, the frozen embedding is the largest single-feature gain in our
ranking-AUC ladder, worth +0.59\% on Discovery and +0.57\% on Search, and an in-model
DIN attention module adds only marginally on top of it. Notably, one embedding helps on surfaces with opposite consumer intent ---
the browse-driven \emph{exploration} surface (Discovery)
and the query-driven \emph{intent} surface (Search). A single daily inference job keeps the
embedding fresh, and it shipped as part of bundled ranker retrains that lifted revenue on both surfaces (+0.79\% Discovery, +0.94\% Search).

\section{Related Work}\label{sec:related}

Sequential recommenders encode a user's interaction history as a token sequence and model it with self-attention to recommend the next item; SASRec~\cite{sasrec} is the canonical causal variant, while BERT4Rec~\cite{bert4rec} uses a bidirectional masked-item (Cloze) objective. A more recent line reframes recommendation as generation: GPT4Rec~\cite{gpt4rec} generates natural-language queries from a user's history and retrieves items with a search engine, TIGER~\cite{tiger} casts retrieval as generating semantic item IDs, and large sequential transducers such as HSTU~\cite{hstu} reformulate both retrieval and ranking as generative sequential transduction at industrial scale. These systems are optimized for the recommendation task itself --- retrieval or ranking --- not for emitting a reusable representation to be consumed by other models. A parallel thread learns user embeddings directly and consumes them as features: YouTube~\cite{youtubednn} trains a user vector against a watch-prediction objective, and Airbnb~\cite{airbnbemb} learns skip-gram listing and user-type embeddings for search ranking. Closest to our setting are reusable industrial user representations consumed by separate downstream models: PinnerFormer~\cite{pinnerformer} trains a user embedding with a dense all-action loss over long-term future engagement, while PinFM~\cite{pinfm} pretrains a billion-scale foundation model and couples it into rankers through candidate-aware cross-attention. We share their reusable-representation goal but differ in objective and loss: rather than long-term all-action prediction or a candidate-conditioned foundation model, we predict a short horizon under an exploration weighting for a repurchase-heavy domain, with a sampled softmax whose hard negatives combine a fixed popular-business pool with businesses shown but not ordered.

Within CTR and ranking models, behavioral sequences are instead consumed by in-model modules: DIN~\cite{din} applies target-aware attention over the behavior history, cross networks (DCN~\cite{dcn}, DCN-V2~\cite{dcnv2}) learn explicit high-order feature interactions, and deployed production rankers increasingly embed a Transformer block over the behavior sequence directly~\cite{bst,transact}. Because such a module lives inside one ranker, its sequence modeling is tied to that model and must be rebuilt for each new ranking task. We take a different route: we train a generative sequential model once and treat it as a frozen, reusable user representation that feeds a separate production ranker, and we find an in-model DIN adds little on top of it.

\section{Method}
\label{sec:method}

\begin{figure*}[t]
\centering
\resizebox{\textwidth}{!}{%
\begin{tikzpicture}[
  font=\small,
  >={Stealth[length=2mm]},
  tok/.style={draw, rounded corners=1pt, minimum size=7mm, inner sep=1pt, fill=black!4},
  gpt/.style={draw, rounded corners=3pt, fill=blue!6, minimum width=30mm, minimum height=22mm, align=center},
  emb/.style={draw, thick, rounded corners=2pt, fill=orange!15, align=center, minimum width=24mm, minimum height=14mm},
  rank/.style={draw, rounded corners=2pt, fill=green!8, align=center, minimum width=22mm, minimum height=14mm},
  surf/.style={draw, rounded corners=2pt, align=center, minimum width=22mm, minimum height=7mm, fill=black!3},
  note/.style={font=\footnotesize, align=center},
]
\node[tok] (s1) {$s_1$};
\node[tok, right=1.4mm of s1] (s2) {$s_2$};
\node[tok, right=1.4mm of s2] (s3) {$s_3$};
\node[tok, right=1.4mm of s3] (sdots) {$\cdots$};
\node[tok, right=1.4mm of sdots] (st) {$s_t$};
\node[note, above=2mm of s3] {consumer store-engagement sequence\\[-1pt]
  {\scriptsize each position: store $+$ business id $+$ 5 context features}};

\node[gpt, right=13mm of st] (gpt)
  {\textbf{Decoder-only}\\\textbf{Transformer}\\[2pt]
   {\scriptsize 3 layers $\cdot$ 3 heads $\times$ 256}\\
   {\scriptsize context 200}};
\draw[->] (st.east) -- (gpt.west);

\node[note, above=7mm of gpt] (obj)
  {predict next few businesses\\[-1pt]
   {\scriptsize sampled softmax over the business vocabulary $+$ hard negatives}\\[-1pt]
   {\scriptsize exploration-weighted loss (up-weight new vs.\ repeat)}};
\draw[->, dashed] (obj.south) -- (gpt.north);

\node[note, below=6mm of gpt] (stage)
  {{\scriptsize Stage 1: click pretrain $\;\longrightarrow\;$ Stage 2: conversion fine-tune (warm start)}};

\node[emb, right=13mm of gpt] (emb)
  {last-token\\hidden state\\[2pt]\textbf{256-d embedding}};
\draw[->] (gpt.east) -- node[note, above, font=\scriptsize] {extract} (emb.west);

\node[rank, right=12mm of emb] (rank) {DCNv2\\ads ranker};
\draw[->] (emb.east) -- node[note, above, font=\scriptsize] {frozen} (rank.west);

\node[surf, right=14mm of rank, yshift=6mm]  (disc)   {Discovery};
\node[surf, right=14mm of rank, yshift=-6mm] (search) {Search};
\draw[->] (rank.east) -- (disc.west);
\draw[->] (rank.east) -- (search.west);
\node[note, above=1.5mm of disc] {\scriptsize ad surfaces};
\end{tikzpicture}%
}
\caption{The generative user representation and how it is reused: a decoder-only
GPT trained on each consumer's store-engagement sequence, whose last-token hidden
state is exported as a frozen 256-d feature and reused by the DCNv2 ranker across
both ad surfaces.}
\Description{A diagram showing a consumer store-engagement sequence feeding into a decoder-only GPT model, whose last-token hidden state is exported as a 256-dimensional embedding and fed frozen into a DCNv2 ranker that serves both the Discovery and Search ad surfaces.}
\label{fig:arch}
\end{figure*}
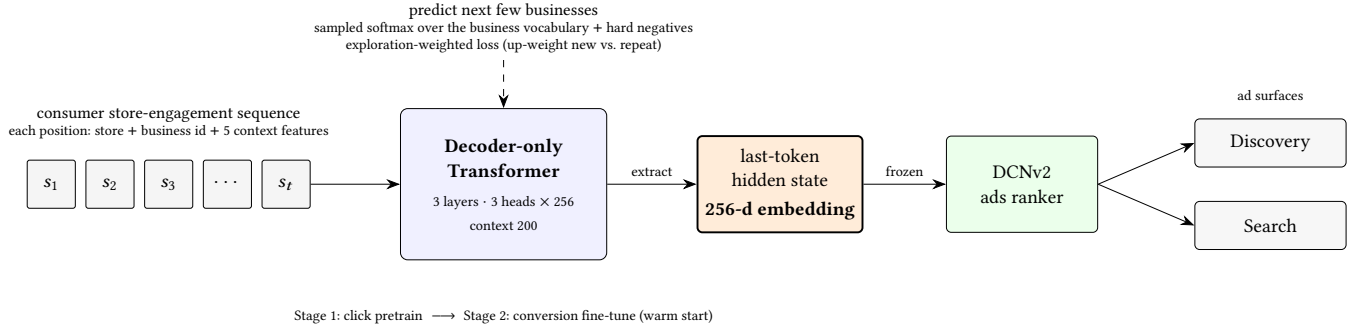

\subsection{Input}
\label{sec:input}
Each entry in a consumer's history is a store they clicked or ordered from,
whether through an ad or organically, identified by a store id and a business id. A \emph{store} is a
single storefront a consumer can order from; a \emph{business} is the merchant
that owns one or more stores, so one business id can span many store
locations. We embed both identifiers with pretrained 256-dimensional lookup
embeddings, together with five learned categorical context features --- the
day-part (breakfast, lunch, dinner, or late night), a weekend flag, local hour,
day of week, and whether the impression was
sponsored --- so each position captures how and when a consumer engaged a
store, not only which one. Ordering these engagements chronologically per
consumer gives the store-engagement sequence, the input to the generative model
(Figure~\ref{fig:arch}).

\subsection{Generative user model}
Given that sequence, we predict the next few businesses the consumer will
engage. At each position the store, business, and context embeddings are
concatenated and projected to the trunk width.

\begin{figure}[t]
\centering
\scalebox{1.0}{%
\begin{tikzpicture}[
  font=\scriptsize,
  >={Stealth[length=1.6mm]},
  node distance=2.4mm,
  fin/.style={draw, rounded corners=1pt, align=center, minimum height=6mm, minimum width=24mm, fill=black!4, inner sep=1.5pt},
  op/.style={draw, rounded corners=1pt, align=center, minimum height=5mm, minimum width=75mm, fill=blue!5, inner sep=2pt},
  blk/.style={draw, rounded corners=2pt, align=center, minimum width=75mm, fill=blue!9, inner sep=2.5pt},
  hid/.style={draw, rounded corners=1pt, align=center, minimum height=5mm, minimum width=75mm, fill=black!7, inner sep=2pt},
  emb/.style={draw, thick, rounded corners=2pt, align=center, fill=orange!16, minimum height=7mm, minimum width=36mm, inner sep=2pt},
]
\node[fin] (bus) {business id emb 256};
\node[fin, left=3mm of bus] (store) {store id emb 256};
\node[fin, right=3mm of bus] (ctx) {5 context, emb 4 each};
\node[op, below=4mm of bus] (proj) {concatenate (532-d) \;$\to$\; project to 768-d trunk \;$\oplus$\; learned positions};
\draw[->] (store.south) -- (store.south |- proj.north);
\draw[->] (bus.south)   -- (bus.south |- proj.north);
\draw[->] (ctx.south)   -- (ctx.south |- proj.north);
\node[blk, below=of proj] (blk) {\textbf{decoder block}\, (pre-norm, causal)\\[1pt]
  masked multi-head self-attention (3 heads $\times$ 256) \;$\to$\; feed-forward network};
\draw[->] (proj) -- (blk);
\node[anchor=west, font=\footnotesize] at ([xshift=1.5mm]blk.east) {$\times\,3$};
\node[hid, below=of blk] (hid) {LayerNorm \;$\to$\; project to 256-d \;$\to$\; per-position hidden states};
\draw[->] (blk) -- (hid);
\node[emb, below=5mm of hid, xshift=-20mm] (out) {\textbf{user embedding}\\last token, 256-d};
\node[emb, below=5mm of hid, xshift=20mm] (trn)
  {\textbf{business embeddings}\\softmax-layer rows, 256-d};
\draw[->] (hid.south) -- (out.north);
\draw[->] (hid.south) -- (trn.north);
\end{tikzpicture}%
}
\caption{Model architecture. Per-position store, business, and
context embeddings are concatenated (532-d), projected to a 768-d trunk, run
through three pre-norm causal decoder blocks, and projected to 256-d. We export
the last-token hidden state as the user embedding and the softmax layer's rows as
business embeddings (\S\ref{sec:extract}).}
\Description{A diagram of the model architecture showing store, business, and context input embeddings concatenated and projected to a 768-dimensional trunk, passed through three pre-norm causal decoder blocks, then normalized and projected to 256 dimensions to produce per-position hidden states, from which the last-token user embedding and the softmax-layer business embeddings are both extracted.}
\label{fig:model}
\end{figure}
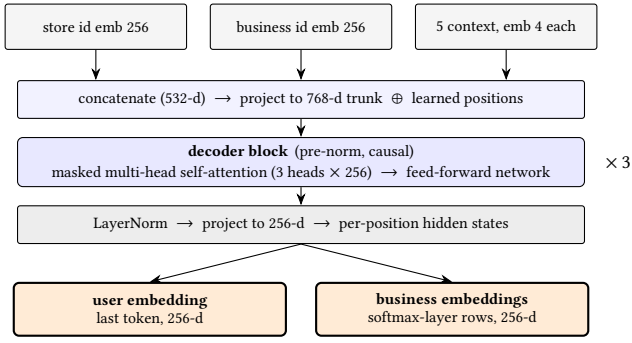

The backbone, detailed in Figure~\ref{fig:model}, is a decoder-only, GPT-style
Transformer with three layers and three attention heads of size 256 --- a
768-wide trunk --- with learned position embeddings over a 200-position context; a final projection yields the
256-dimensional embedding we export. The target at each position is the next
business rather than the individual store --- since one business can span many
store locations, a store-level target would pull the model toward location
instead of merchant preference --- and we also predict a short look-ahead window
of the next few, down-weighting farther horizons. The business
vocabulary runs to hundreds of thousands of entries, and an exact softmax over
that many classes at every position is too expensive at our sequence volume,
so the production model is trained with a sampled softmax --- and how we
\emph{design its negatives} is central to what the representation learns. Each
positive is scored against two kinds of hard negatives --- a fixed pool of about
60K popular businesses, and ones the consumer was shown but did not order ---
together with random and in-batch negatives. The hard negatives force the representation to separate close alternatives,
while random and in-batch draws keep the denominator broad and cheap;
\S\ref{sec:analysis} compares an exact full softmax.

\subsection{Two-stage curriculum and exploration weighting}
We train in two stages. Stage 1 pretrains on click sequences, which are
plentiful and cover a consumer's broad browsing and engagement behavior. Stage 2
fine-tunes on conversion (purchase) sequences to sharpen purchase intent. Rather
than training Stage 2 from scratch, we warm-start it from the Stage-1
checkpoint, so the conversion model keeps the browsing coverage learned from
clicks while specializing toward what consumers actually buy.

Ordering is repurchase-heavy: the next business is usually one already in the
consumer's history. Under a uniform loss the objective is dominated by these
predictable repurchases, and the model can score well without learning anything
about discovery. We therefore up-weight prediction terms whose target is new to the consumer's
history (exploration) and down-weight repeats. This tilts the representation toward discovery, the regime that matters for
surfacing new stores (\S\ref{sec:analysis}).

Putting these pieces together, the training loss is
\begin{equation}
\label{eq:loss}
\mathcal{L}=-\sum_{t}\sum_{n=1}^{N}\alpha_n\,w_{t,n}\,
\log\frac{e^{\,s_{y_{t,n}}}}{e^{\,s_{y_{t,n}}}+\sum_{j\in\mathcal{N}}e^{\,s_{j}}},
\end{equation}
where $s_j=h_t^\top e_j$ scores business $j$ --- its output-layer embedding $e_j$
against the position-$t$ hidden state $h_t$; $y_{t,n}$ is the $n$-step-ahead target
business; $\alpha_n$ weights the prediction horizon rather than applying a time
decay: $\alpha_1{=}1$ and a fixed (not tuned) $\alpha_{n\ge2}{=}0.5$ down-weight
farther-ahead steps; $w_{t,n}{=}0.9$ for a target new to the consumer's history and
$0.1$ for a repeat; and $\mathcal{N}$ is the sampled negative set described above.
We sweep this new-vs.-repeat ratio $w$ (uniform, 3:1, 9:1) in
\S\ref{sec:analysis}; $\alpha_n$ is held fixed.

\subsection{Extraction and reuse}
\label{sec:extract}
The user embedding is the model's last-token hidden state, a single
256-dimensional vector summarizing the consumer's sequence. We export it and
consume it as a frozen dense feature in the downstream DCNv2 ads-ranking model.
No gradient flows back into the generative model. The output layer also holds one 256-d vector per business, scored by dot
product against the hidden state, so each is a business embedding in the same
space as the user embedding; we export both and use them together downstream.

\section{Experimental Setup}
\label{sec:setup}

We run two kinds of offline evaluation. \emph{Intrinsic} evaluation asks how well
the generative model does its own future-business prediction task, scoring it in
isolation. \emph{Extrinsic} evaluation asks the question we actually care about
--- whether the exported embedding makes the downstream ranker better. A model
can win the first and contribute nothing to the second, so we report both. Both use a held-out
set drawn from a period later than the training window, so no metric sees
interactions the model was fit on.

\emph{Intrinsic.} We score the model on
future-converted-business prediction: for each consumer, the businesses they
converted in the two weeks after the training window (with a one-day leakage
buffer). We report Recall@$k$, MRR, and next-item perplexity over a fixed
120K-business candidate set; perplexity is the exponential of the mean next-item
cross-entropy, and lower means more probability on the true business
(\S\ref{sec:analysis}).

\emph{Extrinsic.} We add the user embedding to the production DCNv2
ads ranker and read its Eval AUC on Discovery and Search, as an
incremental ladder that adds components one at a time, so the embedding's gain
is measured on top of everything added before it (Table~\ref{tab:offline}). The embedding is not fed in raw. With $u$ the user embedding and $c$ the candidate
business embedding, the ranker consumes their interaction
\begin{equation}
\label{eq:interaction}
\phi(u,c)=\big[\,u,\; c,\; u\odot c,\; \cos(u,c)\,\big],
\end{equation}
the concatenation of both raw vectors, their element-wise (Hadamard) product
$u\odot c$, and their cosine similarity.

Online evaluation uses randomized A/B tests, with consumers split 50/50 by
\texttt{consumer\_id}. Each test runs for one week, on all
platforms, across the two ad surfaces: Discovery and Search. Every launch ships the user embedding as part of a
bundled ranker retrain rather than as an isolated change, so the measured online
movement is a property of the whole launch and is not attributable to the
embedding alone. We track revenue, CVR, CTR, eCPM, and ROAS
(Table~\ref{tab:online}).
\section{Results}
\label{sec:results}

\subsection{Intrinsic evaluation}
\label{sec:analysis}
Numbers in this subsection are single-run point estimates.
In Table~\ref{tab:ablation}, hard
negatives raise overall recall. A business shown but not ordered is a noisy
negative, since the consumer may still have liked it, so we mix these with
random and in-batch negatives.
Exploration weighting lowers overall recall --- which is dominated by repeat
orders --- but raises recall on businesses the consumer has not engaged before,
monotonically in the ratio: 9:1 loses about 3$\times$ as much overall recall as
3:1 for a smaller further gain on new businesses. Removing repeats from the sequence entirely ---
the \emph{distinct sequence} row --- collapses overall recall, confirming that
repeats carry most of the next-item signal.
We also tried the business's primary category as an input feature; removing it
changed click R@10 by only 0.6 points (46.91\%$\to$46.28\%), so we dropped it
--- the learned business-id embeddings likely already encode it.

\begin{table}[H]
\centering
\small
\caption{Recall ablations (\% change). New R@10: recall on previously unseen businesses (``--'': not recorded). Time-decay
rows use the full-softmax model; full-softmax rows are vs.\ sampled softmax.}
\label{tab:ablation}
\begin{tabular}{@{}lcccc@{}}
\toprule
Design choice & R@10 & R@30 & MRR & New R@10 \\
\midrule
$+$ Hard negatives              & +14.3\%   & +9.0\%    & +8.2\%    & -- \\
Exploration weighting 3:1      & $-$7.0\%  & $-$3.5\%  & $-$5.7\%  & +18.4\% \\
Exploration weighting 9:1      & $-$20.4\% & $-$11.1\% & $-$19.8\% & +29.1\% \\
Distinct sequence               & $-$67.5\% & $-$46.5\% & $-$75.5\% & -- \\
Time-decay attn (replace pos.) & $-$5.0\%  & $-$3.7\%  & $-$5.4\%  & -- \\
Time-decay attn ($+$ pos.)     & $-$3.7\%  & $-$3.1\%  & $-$3.4\%  & -- \\
Full softmax (click)           & +3.5\%   & +3.3\%   & +2.4\%   & -- \\
Full softmax (conversion)      & $-$0.6\% & $-$0.1\% & $-$0.4\% & -- \\
\bottomrule
\end{tabular}
\end{table}

Adding a time-decay bias to attention, so older engagements count less, lowers
recall both when it replaces the learned position embeddings and when it is
added on top of them (Table~\ref{tab:ablation}). What matters is which
businesses a consumer engaged recently, not exactly when.

Finally, as an offline-only comparison, we replace the sampled softmax with an
exact full-softmax over the business vocabulary, made tractable by a
vocabulary-chunked cross-entropy that accumulates the exact partition function
tile by tile without materializing the full logit matrix. It lowers perplexity by 26\% on click and 14\% on
conversion and lifts click recall, but conversion recall stays flat
(Table~\ref{tab:ablation}), so the production model keeps the sampled softmax.

\subsection{Extrinsic evaluation}
\label{sec:offline}
Fed into the ranker, the frozen user embedding is the largest single-feature step in the
offline ladder (Table~\ref{tab:offline}), helping both a browse-driven
exploration surface and a query-driven intent surface. We also tested whether an in-model sequence
module adds anything on top of it: a DIN attention module over the history buys
only a further +0.03\%, so the cheap frozen embedding already captures most of
what an in-model sequence module would contribute. On Search, DIN added without
the embedding gives no gain (+0.00\%), while the embedding alone gives +0.57\%.

We also tested exploration weighting's effect in the ranker: using the embedding from the uniformly weighted model gave no gain
in Discovery Eval AUC,
whereas the 9:1 exploration-weighted embedding gains +0.59\%
(Table~\ref{tab:offline}).

\begin{table}[H]
\centering
\footnotesize
\caption{Discovery ranking-AUC ladder: each row's incremental relative Eval-AUC
change over the rows above; row 1 is vs.\ the retrained production DCNv2.
The uniform and 9:1 rows are alternatives, each measured on its own rather than
stacked. Search: +0.57\% for the user embedding. Single runs.}
\label{tab:offline}
\begin{tabular}{@{}p{0.76\columnwidth}r@{}}
\toprule
Component added & \% impr. \\
\midrule
Store emb.\ $+$ mean-pooled last-$N$ converted-store emb. & +0.33\% \\
$+$ user embedding, uniform weighting & $\approx$0.00\% \\
$+$ user embedding, 9:1 weighting, as $\phi(u,c)$ (Eq.~\ref{eq:interaction}) & \textbf{+0.59\%} \\
$+$ DIN attention over the raw history & +0.03\% \\
\bottomrule
\end{tabular}
\end{table}

\subsection{Online experiments}
The embedding shipped as one component of bundled ranker retrains across two
surfaces --- Discovery first, then Search. Both launches raised revenue and CVR
significantly (Table~\ref{tab:online}); each launch changed more than the
embedding alone.

\begin{table}[H]
\centering
\small
\caption{Online A/B lift by surface (\% change).
\textsuperscript{$\ast\ast\ast$}$p{<}0.001$,
\textsuperscript{$\ast\ast$}$p{<}0.01$, \textsuperscript{$\ast$}$p{<}0.05$;
n.s.\ = not significant.}
\label{tab:online}
\begin{tabular}{@{}lcc@{}}
\toprule
Metric & Discovery & Search \\
\midrule
Revenue    & +0.79\%\textsuperscript{$\ast\ast\ast$}   & +0.94\%\textsuperscript{$\ast\ast\ast$} \\
CVR        & +1.17\%\textsuperscript{$\ast\ast\ast$}   & +1.50\%\textsuperscript{$\ast\ast\ast$} \\
CTR        & $-$0.10\%~(n.s.)                          & +1.38\%\textsuperscript{$\ast\ast\ast$} \\
eCPM       & +0.66\%\textsuperscript{$\ast\ast\ast$}   & +1.56\%\textsuperscript{$\ast\ast\ast$} \\
ROAS       & +0.13\%~(n.s.)                            & $-$0.01\%~(n.s.) \\
\bottomrule
\end{tabular}
\end{table}

\section{Discussion}
\label{sec:discussion}

Because the representation is pretrained once and consumed frozen, a single
daily inference job serves embeddings for tens of millions of consumers to both
Discovery and Search. Sequence modeling thus lives in one shared offline job
rather than in an encoder each surface must train and serve.

Future directions include merging clicks and conversions into one sequence
(the staged curriculum currently balances discovery against repurchase),
longer histories with search queries, harder negatives from first-time orders
that were rated poorly and never reordered, geographic context to down-weight
out-of-area history, and a live-session adapter to capture the within-session
intent a daily frozen embedding misses.

\section{Conclusion}

A single generative model of a consumer's store-engagement sequence brings
per-consumer sequence modeling to ads ranking at DoorDash. Its frozen
last-token embedding captures intent that hand-engineered aggregates and
pooled histories miss, and is the largest single-feature gain in our ranking-AUC
ladder on two surfaces with opposite consumer intent. It shipped in bundled ranker retrains
that lifted revenue and conversion on both surfaces. Pretrained once and consumed
frozen, a generative user representation is a practical building block for
large-scale personalization.

\bibliographystyle{ACM-Reference-Format}
\bibliography{refs}

\appendix
\section*{Generative AI Usage Disclosure}
The authors used generative AI assistants (large language model--based coding and
writing tools) in preparing this work: to assist with portions of the
implementation, to draft and edit the text of this paper, and to help produce its
figures. The authors directed and reviewed all such use, and take full
responsibility for the technical content, experiments, results, and conclusions
presented here.
\end{document}